\documentclass[a4paper,11pt]{article}

\usepackage{jheppub}
\usepackage[T1]{fontenc}
\usepackage{amsfonts}
\usepackage{amsmath}
\usepackage{amssymb,epsf}
\usepackage{latexsym}
\usepackage{graphicx,epsfig}

\title{\boldmath A Light Mediator Relating Neutrino Reactions}

\author[a]{I. Alikhanov} 
\author[b]{and E.~A.~Paschos}

\affiliation[a]{Institute for Nuclear Research of the Russian Academy of Sciences,\\ 117312 Moscow, Russia}
\affiliation[b]{Department of Physics, TU Dortmund,\\ D-44221 Dortmund, Germany}

\emailAdd{ialspbu@gmail.com}
\emailAdd{paschos@physik.uni-dortmund.de}

\abstract{
The extension of the standard model with a multiplicative $U(1)_R$ factor is consistent with a light vector boson. In its simplest realization, only right-handed particles carry charges of the new group. In this model, there is a residual $\tau_{3R}$ symmetry and one new coupling constant  which correlates neutrino interactions. We compute new contributions to  antineutrino--electron scattering and coherent scattering on nuclei, and compare them with the XENON1T result.
}


\begin{document}

\maketitle
\flushbottom

\section{Introduction}
In the standard model, right-handed neutrinos are very inert. They are introduced in the couplings of Higgs particles and are observable in interactions where scalar particles are present. There is also the possibility that they modify relations of the standard model. In fact, this happens in models where right-handed neutrinos are embedded in a larger group. A typical and popular extension is based on the group $SU(2)_L\times U(1)_Y\times U(1)_R$ with fermions carrying charges of the three groups. There are many such extensions~\cite{Appelquist:2002mw,Fayet:1989mq,Dutta:2019fxn,Dutta:2020enk,Lindner:2018kjo,DelleRose:2017xil,Lindner:2020kko,Dev:2021otb,our_e}. They differ in the number of Higgs particles and many of them include new fermions~\cite{Dutta:2019fxn,Dutta:2020enk} identified with dark matter.

The simplest extension, considered here, includes one new gauge boson $X_\mu$ and a singlet scalar particle $\sigma(x)$. It is shown that the model is self-consistent and contains a residual $\tau_{3R}$ symmetry. Related are models based on the same group and an enlarged Higgs sector~\cite{Dutta:2019fxn,Dutta:2020enk,Lindner:2020kko}. These models have additional interactions and fermions identified with dark matter.

Masses for the particles are generated by the standard model Higgs $\phi(x)$ and the new scalar $\sigma(x)$ being charged in $U(1)_R$. Both scalars contribute to the masses of the gauge bosons and mix the neutral bosons. A consequence is the presence of neutral current interactions, which now also include couplings with right-handed neutrinos, proportional to their $U(1)_R$ charges. In addition, the new scalar $\sigma(x)$ generates the mass for the $X_\mu$~boson.

Dirac masses for fermions of the standard model are generated by the vacuum expectation value (VEV) of $\phi(x)$. For fermions with the same charge, only the $SU(2)_L$ contributes to the Dirac mass term producing mass matrices proportional to the couplings of the neutral Higgs mesons. Thus, Dirac mass matrices and neutral Higgs couplings can be diagonalized at the same time, eliminating flavor changing neutral currents. Majorana masses are introduced for right-handed neutrinos by the VEV of $\sigma(x)$. Finally, all fermions appear in the interactions with their specific helicities producing chiral interactions. In explicit calculations, it is helpful to use Weyl spinors for them. 

An additional requirement for an ultraviolet complete theory is the cancellation of triangle anomalies. The relations summarized in Table~\ref{tab_y} are necessary, but not sufficient, conditions for the theory to be ultraviolet finite.
In our case, the left-handed neutrinos do not couple to the $U(1)_R$ term and we select their charges to be zero: $Y_e=Y_{e_L}=0$ and $Y_u=Y_{u_L}=0$. Applying these conditions, a $\tau_{3R}$ symmetry survives for quarks and leptons, as shown in Section~\ref{sec3}.  An explicit classification of the triangle diagrams demonstrates that the theory is ultraviolet finite.




The presence of one coupling constant in the $U(1)_R$ group relates several reactions. The elastic antineutrino--electron scattering still allows a small new contribution. The TEXONO experiment gives data integrated over the electron recoiling energy, but still as a function of the incident energy. Integrating the cross section with the propagator of the mediator brings a logarithmic dependence on the mass of the mediator.  The bound we obtain for the coupling constant  at very low energies is consistent with the value expected when the XENON1T events are produced by neutrinos originating in the Sun.
The third process is the coherent scattering of neutrinos on atomic nuclei.

The content of the article is the following. Sections~\ref{sec2} and~\ref{sec3} describe the model and the cancellation of triangle anomalies. The description concentrates to one generation of fermions and must be repeated for three.  Section~\ref{sec4} shows how right-handed neutrinos couple to the light boson mediator through their $U(1)_R$ charge. Finally, we discuss three neutral current reactions, where low energies are preferable because propagator effects are more important.

\section{The Model\label{sec2}}
\subsection{The Gauge Sector}

The new field $X_\mu$ couples to the standard Higgs doublet $\varphi(x)$ through the covariant~derivative

\begin{equation}
D_{\mu}\varphi(x)=\left(\partial_\mu+ig\frac{\tau_3}{2}\hat W^{3}_\mu+ig'\frac{Y}{2}\hat B_{\mu}+ig_X\frac{Y_\varphi}{2}\hat X_{\mu}\right)\varphi(x). \label{eq:2_8}
\end{equation}
Mass terms for the three bosons are generated when $\varphi(x)$ acquires a vacuum expectation value. An additional scalar Higgs $\sigma(x)$ is also introduced to provide another mass term $(1/4)g_X^2Y'^2v_0^2$ for the $X_\mu$ field with $Y'$ being the charge of $\sigma(x)$ in the $U(1)_R$ group. This brings a non-zero mass for the new gauge boson. After symmetry breaking the square of \newpage\noindent the mass matrix for the fields $\hat W^3_\mu$, $\hat B_\mu$ and $\hat X_\mu$ has the form~\cite{Fayet:1989mq,Appelquist:2002mw,DelleRose:2017xil,Dev:2021otb}

\begin{equation}
M^2=\frac{1}{4}\left(\begin{array}{ccc}
g^2 & -gg' & -gg_XY_\varphi \\
-gg' & g'^2 & g'g_XY_\varphi\\
-gg_XY_\varphi & g'g_XY_\varphi & g_X^2{\tilde{Q}_X}^2\\
\end{array}
\right)\frac{v^2}{2},
\label{eq:2_9}
\end{equation}
where $Y_X=Y_\varphi$ is the hypercharge of the Higgs doublet in $U(1)_R$ and
\begin{equation}
M^2_{33}=\frac{1}{4}g_X^2{\tilde{Q}_X}^2\frac{v^2}{2}=\frac{1}{4}\left(g_X^2Y_\varphi^2+g_X^2Y'^2\frac{v_0^2}{v^2}\right)\frac{v^2}{2}.\label{eq:2_10}
\end{equation}
In this specific extension, the photon remains massless and electromagnetism is not affected.

The diagonalization of the mass matrix is well known, with the electromagnetic field retaining its form in the standard model,

\begin{equation}
A_{\mu}(x)=s_W\hat W^{3}_\mu(x)+c_W\hat B_{\mu}(x)\label{eq:2_11}
\end{equation}
and the charge operator being

\begin{equation}
Q=\frac{\tau_3}{2}+\frac{Y}{2}.\label{eq:2_11_1}
\end{equation}
For the computation of the neutral current, it is useful to express the initial gauge fields in terms of the physical fields $A_\mu$, $Z_\mu$ and $X_\mu$:

\begin{equation}
\hat W^3_\mu=s_WA_{\mu}+c_\alpha c_WZ_\mu+s_\alpha c_WX_\mu,\label{eq:2_12}
\end{equation}

\begin{equation}
\hat B_\mu=c_WA_{\mu}-c_\alpha s_WZ_\mu-s_\alpha s_WX_\mu,\label{eq:2_13}
\end{equation}

\begin{equation}
\hat X_\mu=-s_\alpha Z_{\mu}+c_\alpha X_\mu,\label{eq:2_14}
\end{equation}
where $s_W=\sin\theta_W$, $c_W=\cos\theta_W$ with $\theta_W$ being the Weinberg angle and  $s_\alpha=\sin\alpha$, $c_\alpha=\cos\alpha$ define a new mixing angle $\alpha$ given by~\cite{Fayet:1989mq}

\begin{equation}
c_\alpha^2=\frac{g^2+g'^2}{g^2+g'^2+g_X^2Q_X^2}\approx\frac{g^2+g'^2}{g^2+g'^2+g_X^2Y_\varphi^2}.\label{eq:2_15}
\end{equation}
The diagonalization of the mass matrix~\eqref{eq:2_9}---in the limit $(v_0^2/v^2)\ll1$---provides the relation

\begin{equation}
g_Xc_\alpha Y_\varphi=\frac{gs_\alpha}{c_W}\label{eq:2_16}
\end{equation}
which will be useful later on.

\begin {table}
\begin{center}
\begin{tabular}{ |c|c|c| }
  \hline
& $Y_{SM}$ & $U(1)_R$ \\\hline
   $\nu_L$, $e_L$ & -1 & $Y_e$ \\
	    $u_L$, $d_L$ & 1/3 & $Y_u$ \\
			$N_M$ & 0 & $Y_{N_R}=Y_e+Y_\varphi$ \\
     $e_R$ & -2 & $Y_{e_R}=Y_e-Y_\varphi$ \\
     $u_R$ & 4/3 & $Y_{u_R}=Y_u+Y_\varphi$ \\
     $d_R$ & -2/3 & $Y_{d_R}=Y_u-Y_\varphi$ \\
		 \hline
		 $\varphi$ & 1 & $Y_\varphi$ \\ 
  \hline
\end{tabular}
\caption {$U(1)_R$ charges for leptons and quarks.}\label{tab_y}
\end{center}
\end {table}

\subsection{Fermion Sector}

Fermion masses are generated by Yukawa couplings of the Higgs doublet $H(x)$ and the singlet $\sigma(x)$

\begin{equation}
\mathcal{L}_{Y}=Y^{ij}_e\bar l_{Li}\varphi e_{Rj}+Y^{ij}_\nu\bar l_{Li}\tilde \varphi\nu'_{Rj}+Y^{ij}_\sigma\overline{ {\nu'}_{Ri}^C}\nu'_{Rj}\sigma+h.c. \label{eq:b}
\end{equation}
with $Y_e$, $Y_\nu$ and $Y_\sigma$ being $3\times3$ matrices. 
When the neutral Higgs doublet acquires vacuum expectation value, it produces Dirac masses. Since it is the only scalar contributing to Dirac masses, there are no flavor changing interactions (NFCI).
The two contributions produce a mass matrix of the form

\begin{equation}
\left(\bar\nu_L \,\,\,\overline{ {\nu'}_R^C}\right)\left(
\begin{array}{cc}
0&\tilde m_D\\
\tilde m_D^T&M_R\\
\end{array}
\right)\left(
\begin{array}{c}
\nu^C_L\\
{\nu'}_R\\
\end{array}
\right)\label{eq:2_18}
\end{equation}
It has the structure of the seesaw matrix of type-I. The primed neutrino, $\nu'_R$, will be soon replaced by another state. The Majorana mass matrix is symmetric and is diagonalized by a unitary matrix~\cite{Doi:1980yb, Joshipura:2001ya}.

The invariance of the Dirac mass terms under $U(1)_R$ transformations supply relations among the charges of left- and right-handed leptons and quarks. The sum of charges in each trilinear must be zero, thus giving the relations~\cite{Fayet:1989mq}

\begin{equation}
Y_{u_R}=Y_{u_L}+Y_\varphi, \label{eq:2_23}
\end{equation}

\begin{equation}
Y_{d_R}=Y_{u_L}-Y_\varphi, \label{eq:2_24}
\end{equation}

\begin{equation}
Y_{e_R}=Y_{e_L}-Y_\varphi. \label{eq:2_25}
\end{equation}
In our case, a Dirac mass for the neutrinos is generated from the Yukawa coupling to the right-handed neutrino, giving

\begin{equation}
Y_{N_R}=Y_{e_L}+Y_\varphi. \label{eq:2_26}
\end{equation}
In this manner, there are three independent $U(1)_R$ fermionic charges for each generation summarized in Table~\ref{tab_y}.

In the table, we also included the standard model hypercharges denoted by \linebreak\mbox{$Y_{SM}=2(Q-\tau_3/2).$} We mention that whenever all charges are active, the charge assignments in Table~\ref{tab_y}, together with the relation $Y_e+3Y_u=0$, are sufficient for the cancellation of~anomalies.

The symmetric matrix $M_R$ is diagonalized by a unitary matrix $U_R$~\cite{Joshipura:2001ya}

\begin{equation}
U^*_RM_RU^{\dagger}_R=D=\mathrm{diag}(M_1,M_2,M_3).\label{eq:2_19}
\end{equation}
Now, redefining the right-handed neutrinos $\nu_R=U_R\nu'_R$ and the Dirac matrices $m_D=\tilde m_DU_R$, we obtain

\begin{equation}
\left(\bar\nu_L \,\,\,\overline{ {\nu}_R^C}\right)\left(
\begin{array}{cc}
0&m_D\\
m_D^T&D\\
\end{array}
\right)\left(
\begin{array}{c}
\nu^C_L\\
{\nu}_R\\
\end{array}
\right).\label{eq:2_188}
\end{equation}

Equation \eqref{eq:2_18} may now be used to find the eigenvalues (masses) and eigenfunctions of the mass matrix, because the $D$ matrix is diagonal.
We computed the secular equation for the two-family case. In this case, the entire matrix is a $4\times4$ matrix. A straightforward computation produces a $4^{\text{th}}$ degree polynomial in $\lambda$. Its form has the structure   

\begin{equation}
\lambda f\left(\lambda,m_{Dij},D_i\right)+\mathrm{det}(m_D^Tm_D)=0\label{eq:2_20}
\end{equation}
with $f\left(\lambda,m_{Dij},D_i\right)$ being a cubic polynomial in $\lambda$ with the coefficients being functions of elements of the Dirac matrix $m_{Dij}$ and of $D_i$. Thus, whenever $\mathrm{det}(m_D)=0$, then one neutrino mass is zero, independent of values for the elements of the Majorana matrix.

The eigenvalue equation~\eqref{eq:2_20} has the same structure also in models with more than two families. This follows from the structure of the secular equation

\begin{equation}
\mathrm{det}\left|
\begin{array}{cc}
-\lambda & m_D\\
m_D^T&D-\lambda\\
\end{array}
\right|=0\label{eq:2_20s}
\end{equation}
and the definition of a determinant.

In order to account for the very small masses of neutrinos, it is usually assumed that all elements of the matrix $D$ are very large. This is not always the case, because one eigenvalue of $m_D$ may be very small. For instance, when one eigenvalue of $m_D$ is zero, the determinant of $m_D^Tm_D$ is also zero for any values of the elements in $D$. In fact, it has been shown that in seesaw models of type-I one sterile neutrino can have a mass in the eV range~\cite{Branco:2019avf}. Thus, light sterile neutrinos are possible.

\section{Cancellation of Anomalies\label{sec3}}

In addition to the conditions required for the generation of fermion masses, there are restrictions on the hypercharges from the cancellation of triangle anomalies. In the $U(1)_R$ extension of the standard model there are general conditions~\cite{Appelquist:2002mw}, which are consistent with the conditions in Table~\ref{tab_y}. In our case, the left-handed states of quarks and leptons do not couple to the $U(1)_R$ bosons and we must set $Y_e=Y_u=0$. This reduces the three independent parameters of the table to one and introduces the $\tau_{3R}$ symmetry

\begin{equation}
Y_{e_R}=Y_{d_R}=-Y_{\varphi}\,\,\,\text{and}\,\,\,Y_{N_R}=Y_{u_R}=Y_{\varphi}.\label{eq:3_1}
\end{equation}
In addition, the Majorana couplings are "axial-vector" and in the loops when the momenta of integrations are much larger than the masses, their contributions are equivalent to those for Dirac fermions with a $\gamma^5\gamma^\mu$ vertex.

We classify the diagrams according to the number of external $X_\mu$ fields.

\begin{figure}
\centering
\includegraphics[width=1.15\textwidth]{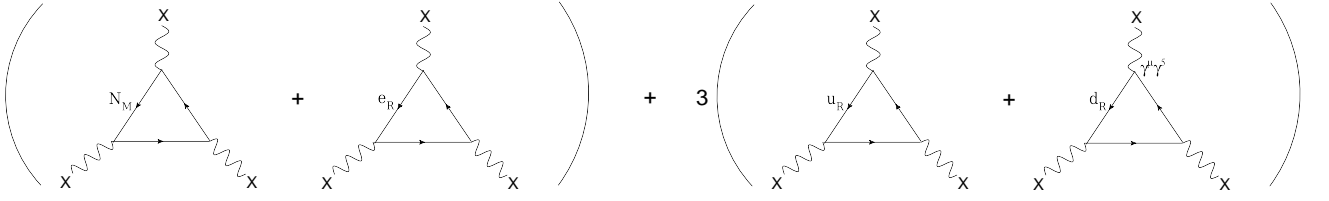}
\caption{Triangle diagrams with three $X_\mu$ external fields.}
\label{fig:1}
\end{figure}

(1) Diagrams with three external fields are shown in Figure~\ref{fig:1}. They cancel by virtue of the condition

\begin{equation}
Y_{N_R}^3+Y_{e_R}^3=(-Y_{\varphi})^3+Y_\varphi^3=0\,\,\,\,\text{and}\,\,\,\,Y_{u_R}^3+Y_{d_R}^3=0.\label{eq:3_2}
\end{equation}

\begin{figure}
\centering
\includegraphics[width=0.75\textwidth]{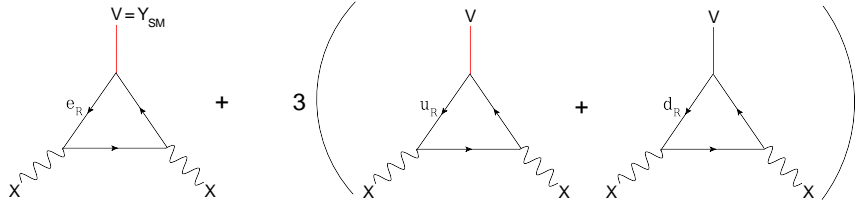}
\caption{Triangle diagrams for the case $Y_{SM}X^2$.}
\label{fig:2}
\end{figure}

(2) For the case $Y_{SM}X^2$, the cancellation of anomalies requires the condition

\begin{equation}
V_{e_R}Y_{e_R}^2+3(V_{u_R}Y_{u_R}^2+V_{d_R}Y_{d_R}^2)=-2Y_{e_R}^2+3\left(\frac{4}{3}Y_{u_R}^2-\frac{2}{3}Y_{d_R}^2\right)=0\label{eq:3_3}
\end{equation}
as it follows from Figure~\ref{fig:2}.

\begin{figure}
\centering
\includegraphics[width=0.75\textwidth]{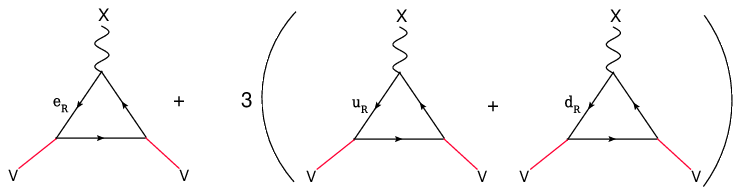}
\caption{Triangle diagrams for the case $Y_{SM}^2X$.}
\label{fig:3}
\end{figure}

(3) For $Y_{SM}^2X$, the conditions required for the cancellation are

\begin{equation}
V_{e_R}^2Y_{e_R}+3(V_{u_R}^2Y_{u_R}+V_{d_R}^2Y_{d_R})=4Y_{e_R}+\left(\frac{16}{3}Y_{u_R}+\frac{4}{3}Y_{d_R}\right)=0.\label{eq:3_4}
\end{equation}
The corresponding diagrams are shown in Figure~\ref{fig:3}.

\newpage (4) For $(QCD)^2X$, the cancellations require

\begin{equation}
Y_{u_R}+Y_{d_R}=0.\label{eq:3_5}
\end{equation}

\begin{figure}
\includegraphics[width=1.15\textwidth]{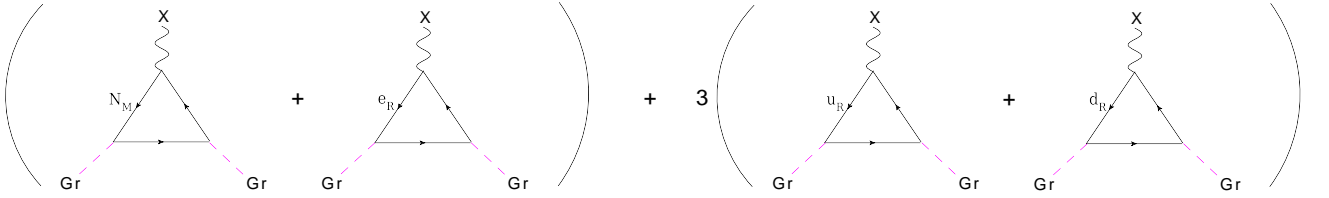}
\caption{Triangle diagrams for the case $(Gravity)^2X$.}
\label{fig:4}
\end{figure}

(5) For $(Gravity)^2X$, the diagrams in Figure~\ref{fig:4} lead to the condition

\begin{equation}
Y_{N_R}+Y_{e_R}+3\left(Y_{u_R}+Y_{d_R}\right)=0.\label{eq:3_6}
\end{equation}

As we mentioned earlier, the charge assignments in Table~\ref{tab_y} with the condition \linebreak\mbox{$Y_e+3Y_u=0$} are sufficient for the cancellation of anomalies. Our approach satisfies this requirement with additional condition $Y_e=Y_u=0$.

\section{Structure of the Neutral Currents\label{sec4}}

Neutral current interactions for left-handed and right-handed quarks and leptons are prescribed by the Lagrangian

\begin{equation}
\mathcal{L}_{NC}=\bar\Psi_Li\gamma^\mu\left[ig\frac{\tau_3}{2}\hat W^{3}_\mu+ig'\frac{Y}{2}\hat B_{\mu}\right]\Psi_L+\bar\Psi_Ri\gamma^\mu\left[ig'\frac{Y}{2}\hat B_{\mu}+ig_X\frac{Y_\Psi}{2}\hat X_{\mu}\right]\Psi_R. \label{eq:4_1}
\end{equation}
The hats on the fields denote flavor fields that are replaced, with the help of~\eqref{eq:2_12}--\eqref{eq:2_14}, by physical gauge fields. The complete neutral current interaction is now written in matrix~form:

\begin{equation}
\left(
\begin{array}{c}
\mathcal{L}_{NC}(Z)\\
\mathcal{L}_{NC}(X)\\
\end{array}
\right)=-\left(
\begin{array}{cc}
Z_\mu\frac{gc_\alpha}{c_W} &\,\, -Z_\mu g_Xs_\alpha\\
X_\mu\frac{gs_\alpha}{c_W} &\,\, X_\mu g_Xc_\alpha\\
\end{array}
\right)\left(
\begin{array}{c}
\bar\Psi_L\gamma^\mu\frac{\tau_3}{2}\Psi_L-s_W^2\sum\limits_\Psi \bar\Psi\gamma^\mu Q\Psi \\
\sum\limits_{\Psi_R} \bar\Psi_R\gamma^\mu\frac{Y_\Psi}{2}\Psi_R\\
\end{array}
\right).\label{eq:4_2}
\end{equation}
The upper component in the column matrix on the right-hand side has the structure of the standard model and the lower component is the new contribution from the initial $\hat X_\mu$ field~\cite{DelleRose:2017xil}. The $2\times2$ matrix emphasizes the mixing between the gauge bosons. We notice that the two contributions for the $X_\mu$ couplings are related through~\eqref{eq:2_16} and are of the same magnitude. For the first generation of quarks, the left-handed state is $\Psi_L^T=(u_L\,\,d_L)$ and the right-handed states are $u_R$ and $d_R$. The summation $\sum\limits_\Psi$ is over the charged states and $\sum\limits_{\Psi_R}$ runs over $u_R$ and $d_R$ carrying the $U(1)_R$ charges $Y_\varphi$ and $-Y_\varphi$, respectively. Introducing them in~\eqref{eq:4_2}, we obtain the $\mathcal{L}_{NC}(X)$ interaction for quarks

\begin{eqnarray}
-\mathcal{L}_{NC}(X)=X_\mu\frac{g_Xc_\alpha}{4}Y_\varphi\left[\bar u\gamma^{\mu}(1-\gamma^5)u-\bar d\gamma^{\mu}(1-\gamma^5)d-s_W^2\left(\frac{8}{3}\bar u\gamma^\mu u-\frac{4}{3}\bar d\gamma^\mu d\right)\right]\nonumber\\+X_\mu\frac{g_Xc_\alpha}{4}Y_\varphi\left[\bar u\gamma^{\mu}(1+\gamma^5)u-\bar d\gamma^{\mu}(1+\gamma^5)d\right]\nonumber\\
=X_\mu\frac{g_Xc_\alpha}{4}Y_\varphi\left[2(\bar u\gamma^{\mu}u-\bar d\gamma^{\mu}d)-s_W^2\left(\frac{8}{3}\bar u\gamma^\mu u-\frac{4}{3}\bar d\gamma^\mu d\right)\right].\label{eq:4_3}
\end{eqnarray}
The final couplings are vector in ordinary space and their isospin content is mostly isovector.

For the leptons, we use left-handed states, and for the right-handed states, we introduce the right-handed electron and the Majorana neutrino, obtaining the couplings to $X_\mu$:

\begin{eqnarray}
-\mathcal{L}^{\text{leptons}}_{NC}(X)=X_\mu\frac{g_Xc_\alpha}{4}Y_\varphi\left[\bar u_\nu\gamma^{\mu}(1-\gamma^5)u_\nu-\bar u_e\gamma^{\mu}(1-\gamma^5)u_e+4s_W^2\bar u_e\gamma^\mu u_e\right.\nonumber\\\left.+(\bar u_N\gamma^\mu\gamma^5u_N-\bar u_e\gamma^\mu(1+\gamma^5)u_e)\right]\nonumber\\
=X_\mu\frac{g_Xc_\alpha}{4}Y_\varphi\left[\bar u_N\gamma^{\mu}\gamma^5u_N+\bar u_\nu\gamma^{\mu}(1-\gamma^5)u_\nu-2(1-2s_W^2)\bar u_e\gamma^{\mu}u_e\right].\label{eq:4_4}
\end{eqnarray}

There are two "axial-vector" terms, one for the Majorana particle and the other for the normal neutrino. When both of them are present in a beam, they contribute incoherently. The term $\bar u_\nu\gamma^{\mu}(1-\gamma^5)u_\nu$ is for the traditional neutrino and is generated from the mixing of $X_\mu$ with the $Z_\mu$ boson.

\section{A Bound for the Mixing Angle\label{sec5}}
The model is subject to constrains. The eigenvalues of the mass matrix~\eqref{eq:2_9} are complicated algebraic functions, but the leading terms are simple and sufficient for the analysis. The $Z$ boson mass has a new contribution 

\begin{equation}
m_Z=\frac{1}{2}v\sqrt{g^2+g'^2+g_X^2Y_\varphi^2}.
\end{equation}
A limit on the mixing angle is obtained from the $\rho$ parameter, which is modified

\begin{equation}
\frac{m_W^2}{m_Z^2}=\frac{g^2}{g^2+g'^2+g_X^2Y_\varphi^2}=c_W^2c_\alpha^2.\label{eq:23n}
\end{equation}
The experimental value for the ratio 
\begin{equation}
\rho=\frac{m_W^2}{m_Z^2c_W^2}=c_\alpha^2=1.00037\pm0.00023
\end{equation}
restricts the mixing angle to within two standard deviations. 
\begin{equation}
0.99991\leq c_\alpha^2\leq1\label{eq:5_4}
\end{equation}
restricting $s_\alpha^2<8\times10^{-5}$ or $s_\alpha\leq9\times10^{-3}\approx0.01$. The value of $c_\alpha^2$ is very close to unity so that the strengths of the neutral relative to the charged current coupling are for practical purposes the same.

Combining~\eqref{eq:5_4} with the relation~\eqref{eq:2_16} leads to the upper bound
\begin{equation}
g_Xc_\alpha Y_\varphi<2.3\times10^{-3}.\label{eq:restr_c}
\end{equation}
We show later on that the upper bound produces substantial effects on antineutrino--electron scattering and must be restricted even more.

\section{Experimental Bounds\label{sec6}}

The fact that the model depends on a single new parameter $(g_Xc_\alpha/4)Y_\varphi$ relates several processes. We discuss three reactions that are accessible to running experiments.

\subsection{Elastic Neutrino--Electron Scattering\label{subsec6}}

A new contribution arises in elastic scattering of neutrinos or antineutrinos on electrons mediated by the exchange of the $X_\mu$ boson. The amplitude for the process includes the $X_\mu$ propagator with $T_e$ the kinetic energy of the recoiling electron

\begin{eqnarray}
\mathcal{M}_e=i\left(\frac{g_Xc_\alpha}{4}\right)^2Y_\varphi^2\frac{2(1-2s_W^2)}{m_X^2+2T_em_e}\bar u_\nu\gamma^\mu(1-\gamma^5) u_\nu\bar u_e\gamma_\mu u_e.\label{eq:6_1}
\end{eqnarray}
The amplitude modifies the vector coupling to electrons. Several experiments measured the process~\cite{Vilain:1993kd,Deniz:2009mu,Park:2015eqa} including the energy spectrum of the recoiling electrons.  The TEXONO collaboration investigated the region of lower energies $3<E_\nu<8$ MeV and reported the results integrated over the recoiling energy of the electrons~\cite{Deniz:2009mu}. The error bars for the integrated cross section are $25\%$.  We demand that the interference of the new amplitude with the amplitude from the standard model is less than~20~\%.

\begin{figure}
\centering
\includegraphics[width=0.8\textwidth]{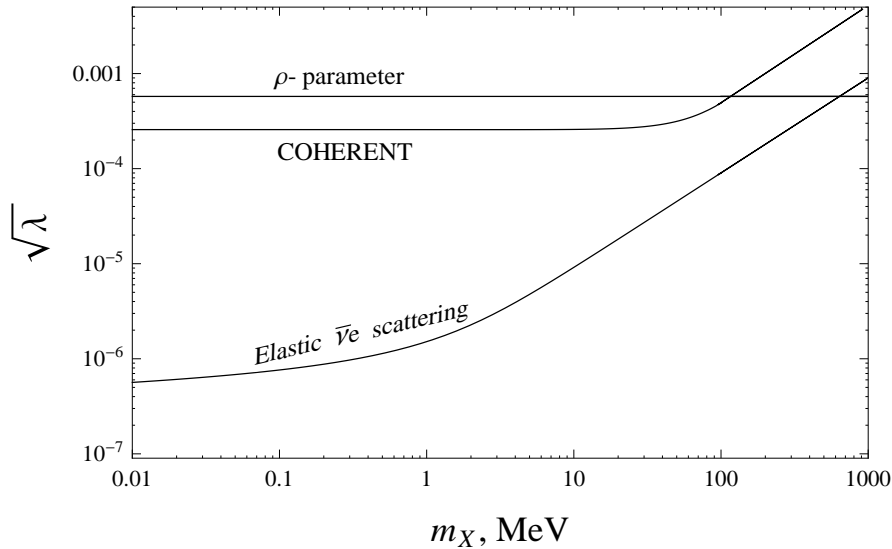}
\caption{Upper limits for the coupling constant as functions of $m_X$ obtained from various experiments. The upper curve is from the $\rho$-parameter; middle curve from COHERENT~\cite{Akimov:2017ade} and lower curve from elastic antineutrino--electron scattering~\cite{Deniz:2009mu}.}
\label{fig:26}
\end{figure}

This gives the condition

\begin{equation}
\frac{\dfrac{\sqrt{2}}{G}\lambda\int_0^{1}[(g_V-g_A)+(g_V+g_A)(1-x)^2]\dfrac{1}{m_X^2+sx}dx}{(g_V-g_A)^2+\frac{1}{3}(g_V+g_A)^2}\leq0.1.
\end{equation}
In the relation we kept only the interference term and integrated over $x$. For $m_X^2\leq s$, we use the exact integral, and for $m_X^2\gg s$, we approximate the propagator by $1/m_X^2$. A consequence of the condition is the upper bound for the coupling constant as a function of $m_X$.
The bound for the coupling constant

\begin{equation}
\sqrt{\lambda}=\frac{g_Xc_\alpha}{4}Y_\varphi\label{copl_s}
\end{equation}
as a function of $m_X$ and for $E_\nu=8$~MeV is shown in Figure~\ref{fig:26}, denoted by elastic $\bar\nu e$ scattering. It is important to keep the mass $m_X$ because it defines the contribution for the lower limit of integration. We note that the dependence of the coupling constant on the mass of the mediator for $m_X^2\ll2m_eT_e$ is smoother without a sharp break. For the TEXONO experiment, we take $2m_eT_e=7$~MeV$^2$, thus for $m_X\ll1$~MeV the bound for the coupling constant is almost independent of $m_X$. 

The recoiling spectrum of the electrons will remain an important channel for investigations beyond the standard model~\cite{Ge:2017mcq}; for this reason there are updates for the standard model parameters~\cite{Canas:2016vxp} and also extensive studies of radiative corrections~\cite{Tomalak:2019ibg}.

The XENON1T experiment~\cite{Aprile:2020tmw} reported an excess of events at recoiling energies $2.5~\text{keV}<T_e<30~\text{keV}$.
The origin of the excess has been interpreted in various ways. One of them attributes the excess to the scattering of solar neutrinos on atomic electrons~\mbox{\cite{Lindner:2020kko,Boehm:2020ltd,AristizabalSierra:2020edu,Khan:2020vaf}}. Since the events are at very low energies they can be produced by the light mediator discussed in this article. Thus we extended our calculation~\cite{our_e} of the neutrino--electron scattering in Figure~\ref{fig:26} down to 10~keV with an upper limit for the coupling constant $\sqrt{\lambda}\leq8\times10^{-7}$. This is in agreement with the range $8\times10^{-7}<\sqrt{g_eg_V}<10^{-6}$ given in the articles~\cite{Lindner:2020kko,Boehm:2020ltd,AristizabalSierra:2020edu,Khan:2020vaf} for $10~\text{keV}\leq m_X\leq 100~\text{keV}$.

\subsection{Coherent Elastic Neutrino Scattering on Nuclei\label{sec7}}
Many years after the original suggestion~\cite{Freedman:1973yd}, coherent elastic scattering of neutrinos (CE$\nu$NS) on atomic nucleus was observed~\cite{Akimov:2017ade}. The process is sensitive to new interactions and to the value of the Weinberg angle~\cite{Scholberg:2015}. A new measurement for the so-called quenching factor of CsI(Na)~\cite{Collar:2019ihs} reduced the experimental error and motivated theoretical studies~\cite{Papoulias:2019txv,Khan:2019cvi,Giunti:2019xpr} that reanalyzed the data and obtained the value $\sin^2{\theta_W}=0.238\pm0.045$ for the Weinberg angle. In the previous section, we used this value for determining the vector coupling $g_V$.

The dominant term for coherent scattering of neutrinos on atomic nuclei is proportional to the baryonic current, which contributes coherently over the entire nucleus. The new amplitude is

\begin{equation}
\mathcal{M}_C=i\frac{2}{3}\frac{\lambda s_W^2}{m_X^2+2MT}\bar u_\nu\gamma_{\mu}(1-\gamma^5)u_\nu\left[\bar u\gamma^\mu u+\bar d\gamma^\mu d\right].\label{eq:34n}
\end{equation}
This amplitude contributes coherently to the cross section, producing

\begin{equation}
\frac{d\sigma}{dT}=\left[1+\frac{4\sqrt{2}}{3G}\frac{A}{Q_W}\frac{\lambda s_W^2}{m_X^2+2MT}\right]^2\left(\frac{d\sigma}{dT}\right)_{\text{SM}}. \label{eq:33n}
\end{equation}
Here, $(d\sigma/dT)_{\text{SM}}$ is the standard model cross section with  the recoiling kinetic energy of the nucleus~$T$, $Q_W=[N-(1-4s_W^2)Z]$ and $A$ is the atomic number of the nucleus. The coupling constant $\lambda$ is the same as in~\eqref{copl_s}. The neutrinos in the experiment~\cite{Akimov:2017ade,Scholberg:2015} are in the energy range $16~\text{MeV}<E_\nu<53~\text{MeV}$ and we will use the median value $E_\nu=30$~MeV, which gives

\begin{equation}
2MT=4E_\nu^2=3.6\times10^{3}~\text{MeV}^2.\label{q:37n}
\end{equation}
Allowing $1\sigma$ uncertainty in the Weinberg angle brings a change on the value of the cross section larger than $20\%$. Thus, assuming the contribution from the new amplitude to be smaller than $10\%$ of the amplitude from the standard model restricts the coupling constant to be below the curve denoted as COHERENT in Figure~\ref{fig:26}. Non-standard neutrino couplings to a light vector mediator have been studied by other groups~\cite{Kosmas:2017tsq,Billard:2018jnl,Lindner:2016wff,Abdullah:2018ykz} and limits were obtained for phenomenological couplings. Their results are comparable to our bounds in Figure~\ref{fig:26}. The limits from CE$\nu$NS are less stringent than those coming from $\nu-e$ scattering; however, they are new and there is a lot of room for significant improvements~\mbox{\cite{Kosmas:2017tsq,Billard:2018jnl}}.

\section{Summary\label{conclusion}}
Neutrinos produced in laboratories or coming from astrophysical sources induce reactions compatible with a light mediator. There are also many theoretical models where a mediator is present. A popular family of models is based on the group  $SU(2)_L\times U(1)_Y\times U(1)_R$. These minimal extensions have an additional gauge boson as the mediator and one or more scalar particles.

One such model was described in Sections~\ref{sec2} and~\ref{sec3} of the article borrowing results from earlier publications~\cite{Appelquist:2002mw,Dutta:2019fxn,Lindner:2018kjo,Lindner:2020kko,Dev:2021otb}, especially~\cite{our_e}. The model has a new gauge boson, $X_\mu$, which mixes with the standard $Z_\mu$. We selected a special assignment of quantum numbers: all left-handed fermions are singlet in $U(1)_{R}$ and only right-handed fermions carry $U(1)_R$ charges. As a result, there is only one new coupling constant (Equation~(\ref{copl_s})). Thus, several reactions are related to each other, producing a fertile ground for investigations. It is also natural to generate a mass matrix for neutrinos with a see-saw mechanism of type-I. Furthermore, the mixing of the gauge bosons produces a neutral current whose left-handed components have the structure of the standard model, plus a new component from right-handed states, including right-handed neutrinos (Equation~(\ref{eq:4_2})). When the two components (left-handed and right-handed) refer to the same fermion, they add up and produce a vector interaction.

Several neutral current reactions are modified due to the exchange of the light mediator. Elastic neutrino--electron or antineutrino--electron scattering experiments have an accuracy of~25~\%, which provides an upper bound for the coupling constant as a function of $m_X$. The new interaction contributes to the vector coupling of the electron and modifies the recoiling spectrum of electrons~\cite{our_e}. It has been proposed that the XENON1T signal may be attributed to solar neutrinos from the $p-p$ process~\cite{Lindner:2020kko,Boehm:2020ltd,AristizabalSierra:2020edu,Khan:2020vaf}. In this case, there is a value for the coupling from the $10~\text{keV}<m_X<100~\text{keV}$ region. Adopting this value for the coupling constant makes the cross sections at recoiling energies of MeV very small.

A third process is coherent scattering of neutrinos on atomic nuclei~\cite{Akimov:2017ade,Scholberg:2015}. The corresponding amplitude has a baryonic current coupled to quarks, which enhances the cross section. The dependence of the cross section on the small recoiling energy is the same as in the standard model, and the signature will be a change in the magnitude of the cross Section \cite{Kosmas:2017tsq,Billard:2018jnl,Lindner:2016wff,Abdullah:2018ykz}.

 This work is an improvement of the reference~\cite{our_e} [arXiv:1902.09950]   indicating that

\noindent (i) chiral gauge invariance is maintained whenever the initial fermions are chiral, and
 
\noindent (ii)  the relevance of the XENON1T  result.


\acknowledgments
One of us (I.A.) was partly supported by the Program of fundamental scientific research of the Presidium of the Russian Academy of Sciences "Physics of fundamental interactions and nuclear technologies".

\end{document}